\documentclass{article}
\usepackage{graphicx} % Required for inserting 
\usepackage[utf8]{inputenc}
\usepackage{hyperref}

\usepackage{float}
\usepackage{multicol}
\usepackage{amsmath}

\usepackage{mathrsfs}
\usepackage{amssymb}
\usepackage{color}
\usepackage{lipsum}\usepackage{fontsize}
  \changefontsize[10.0pt]{8pt}

\usepackage[total={6.5in, 9in}]{geometry}
\usepackage{titling}
\newcommand{\SM}[1]{\textcolor{black}{#1}}

\usepackage[super,numbers,sort&compress]{natbib}
\title{Imperfect Turing Patterns: Diffusiophoretic Assembly of Hard Spheres via Reaction-Diffusion Instabilities}
\author{Siamak Mirfendereski$^1$, and Ankur Gupta$^1$$^*$}
\date{%
    $^1$ \small Department of Chemical and Biological Engineering, University of Colorado Boulder, Boulder, CO 80309, USA\\%
    $^*$ Correspondance: ankur.gupta@colorado.edu\\[2ex]%
}

\begin{document}

\maketitle

\begin{abstract}    
%\section*{Summary}
Turing patterns are stationary, wave-like structures that emerge from the nonequilibrium assembly of reactive and diffusive components. While they are foundational in biophysics, their classical formulation relies on a single characteristic length scale that balances reaction and diffusion, making them overly simplistic for describing biological patterns, which often exhibit multi-scale structures, grain-like textures, and inherent imperfections. Here, we integrate diffusiophoretically-assisted assembly of finite-sized cells, driven by a background chemical gradient in a Turing pattern, while also incorporating intercellular interactions. This framework introduces key control parameters, such as the P\'{e}clet number, cell size distribution, and intercellular interactions, enabling us to reproduce strikingly similar structural features observed in natural patterns. We report imperfections, including spatial variations in pattern thickness, packing limits, and pattern breakups. Our model not only deepens our understanding but also opens a new line of inquiry into imperfect Turing patterns that deviate from the classical formulation in significant ways.
\end{abstract}

\begin{multicols*}{2}

\setlength{\emergencystretch}{3em}

\section*{Introduction}
 Nonequilibrium assembly is critical for theoretical biophysics~\cite{fang2019,mann2009} as it underpins morphogenesis \cite{turing1952}, pattern formation on vertebrate skin \cite{kratochwil2023}, metastatic invasion and phenotype switching \cite{jiao2011}, among others. Reaction-diffusion (RD) instabilities \cite{kondo2010,kratochwil2023} are used to describe the nonequilibrium assembly. The interplay of diffusive and reactive components can give rise to the formation of nonlinear wavelike patterns. Typically, the characteristics of the pattern are determined by the diffusion and reaction rates between the components \cite{kondo2010}. Turing proposed such an RD description in his seminal work \cite{turing1952}. Experimental evidence supports the presence of Turing patterns in processes such as the formation of human fingerprint ridges \cite{glover2023}, the spatial arrangement of the hair follicle \citep{sick2006}, finger formation during early embryo development \cite{raspopovic2014}, and zebrafish embryogenesis \cite{muller2012}.
Following Turing's work, different mathematical models have been proposed as refinements to the original model. For instance, the Gierer-Meinhardt, an activator-inhibitor model, \cite{gierer1972} reproduces the wave-like patterns on Angelfish \cite{kondo1995} and the color patterns on seashells \cite{meinhardt2009}. Similarly, the Brusselator, an activator-substrate model \cite{prigogine1967,prigogine1968}, reproduces hexagonal and strip patterns of an ornate boxfish and the double dot patterns on a jewel moray eel \cite{alessio2023}.  More recently, a cell-cell interaction  model \cite{nakamasu2009,volkening2018} has been proposed for strip-like patterns on zebrafish as well as their adapting behavior after being locally subjected to ablation. Clearly, RD description is a powerful tool to replicate patterns observed in biological systems \cite{turing1952,gierer1972,prigogine1967,meinhardt2009,prigogine1968,kondo1995,kondo2010}. However, some crucial features are required for patterns to emerge. 

\begin{figure*}[htb]
    \centering
\includegraphics[width=0.98\textwidth]{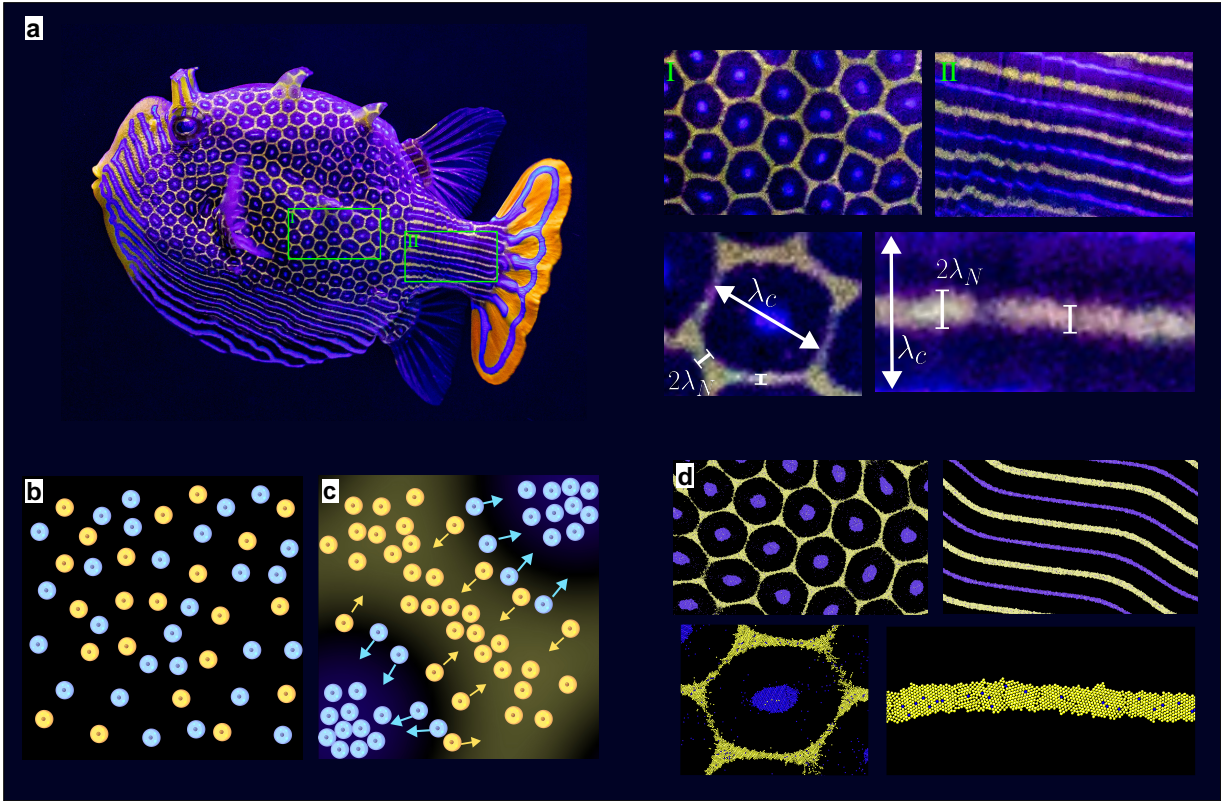}    \caption{ \textbf{Natural patterns in fish and proposed framework for replicating these patterns.} (a) Skin patterns of a male ornate boxfish, showcasing distinct hexagonal and striped structures. \SM{Magnified views of the hexagonal and striped patterns on the fish skin are shown on the right.} These patterns are characterized by two length scales: the pattern size ($\lambda_C$) and the pattern thickness ($\lambda_N$). The patterns exhibit deviations from perfect geometries, featuring imperfections and defects alongside spatial variations in thickness. \SM{Additionally, these patterns possess ``grain-like" features.}
(b,c) Schematic of the proposed simulation framework:  \SM{Diffusiophoretic assembly of cells in response} to the reaction-diffusion of biochemical molecules in the background. (b) The simulation starts by a random distribution of two or more types of cells, with different mobilities and sizes. (c) The solute concentrations are determined through continuum reaction-diffusion descriptions \SM{(equation (1))}, on top of which the cells assemble via diffusiophoresis and form  Turing patterns. The interaction between cells is included via a hard-sphere potential.
(d) Simulated hexagon and stripe patterns \SM{obtained by diffusiophoretic assembly of two types of cells on top of chemical patterns generated by using} the Brusselator model \cite{prigogine1967, pena2001}. The simulations effectively reproduce the natural skin patterns, including the grain-like feature and imperfections. Details of the simulation method and parameters are provided in the Methods section. Photo courtesy of the Birch Aquarium at the Scripps Institution of Oceanography.
} 
    \label{fig:compare1}
\end{figure*}

The key requirements of an RD system that result in the formation of patterns are short-range self-enhancement and long-range inhibition~\cite{koche1994}. The property of self-enhancement refers to the autocatalytic nature of a component. In contrast, the long-range inhibition refers to an antagonistic component that diffuses fast to contain the self-enhancing component. By carefully defining the reaction mechanisms that fulfill the aforementioned criteria, one can create a wide range of patterns ~\cite{sanderson2006,koche1994}. Nonetheless, some critical shortcomings of RD descriptions exist. 
\par{} The length scales in RD descriptions are given only by the balance of reaction and diffusion. For instance, in a stripe-like pattern, the spacing between stripes ($\lambda_C$) and thickness of the stripes ($\lambda_N$) are the same length scale. Similarly, in a hexagonal pattern, the edge length of a hexagon and the thickness of the hexagon are equal. This contrasts with the patterns observed in nature. For example, in an ornate boxfish, there are two length scales for the hexagonal and stripe patterns (see Fig.~\ref{fig:compare1}a). Second, the patterns that emerge from the RD descriptions are smooth and continuous, in contrast to the ones observed in real biological systems where the discontinuities are common. Moreover, the natural patterns possess  ``grain-like" features, i.e., the small-scale roughness in patterns that arises due to the discrete nature of cells, which are not observed in continuum simulations (Fig.~\ref{fig:compare1}a). 
\par{} There have been attempts to resolve the challenges within the RD framework. We recently reported that  diffusiophoresis ~\cite{alessio2023} --  the motion of colloidal particles driven by chemical concentration gradients \cite{shin2016,shi2016,wilson2020,gupta2020,velegol2016,shim2022,ganguly2023, shim2022diffusiophoresis} -- of cells creates a self-steepening effect where $\lambda_N$ and $\lambda_C$ become decoupled such that $\frac{\lambda_N}{\lambda_C}$ becomes a function of  P\`{e}clet number ($Pe$) of the cells, where $Pe$ represents the ratio of advection to diffusion of cells. The basic idea is that  cells are diffusiophoretic \cite{fuji2000, liu2023},  ranging in size from 1 to 30 $\mu$m \citep{velegol2016, shim2022, liu2023}, and are embedded in a medium with concentration gradients of biochemical signals, such as that of morphogens, generated by biological reactions. The biochemical signals provide the basis for their diffusiophoretic responses and consequent assembly of cells. Through the inclusion of diffusiophoresis, the pattern type and the pattern size (i.e., $\lambda_C$) are set by the balance of reaction and diffusion, just like classical Turing models. However, the pattern thickness  (i.e., $\lambda_N$) is set by Pe. The higher the P\`{e}clet number, the finer the thickness of the patterns, thus revealing patterns sharper than those predicted by the RD descriptions and bringing them closer to natural patterns.
\par{} While inclusion of diffusiophoresis  can enable the simulation of sharper patterns,  the grain-like features and imperfect structures remain elusive in the existing continuum framework. To this end, recent work has suggested that color changes~\cite{manukyan2017, fofonjka2021reaction} on lizard skin scales can be simulated through a cellular automaton. This approach proposed that the color of a scale changes based on rules that are derived from an RD framework. This can replicate the dynamics of color changes in scales observed on lizard skin over several years. This framework is compelling since it sheds light on the discreteness of colors. However, the scales themselves are stationary and can only change their state (i.e., color).  This knowledge gap -- that continuum models are unable to capture the discreteness of natural pattern formation while the discrete models are unable to capture the assembly -- motivates this work.
\par{} Our idea is to represent assembling entities, which we refer to as cells, as hard spheres that are assembling diffusiophoretically within a background of biochemical species such that biochemical species are governed by a continuum RD description. Over this background, such that each cell moves  (Fig.~\ref{fig:compare1}b) via diffusiophoresis and Brownian motion, enabling us to introduce a natural way for the cells to assemble on top of a chemical pattern (Fig.~\ref{fig:compare1}c). Due to hard sphere interactions, the cells do not overlap, enabling us to capture the discreteness of the system and the effect of particle size distribution. Supplementary Movies S1 and S2 from the current simulations show diffusiophoretic motion and assembly of finite-sized cells in response to chemical gradients, resulting in the formation of hexagonal and stripe patterns, respectively. We can also simulate polydisperse particle suspensions that show trapped particles and grain-like features; see Supplementary Movies S3 and S4. The simulated hexagons and stripes appear distinctively similar to natural patterns (Fig.~\ref{fig:compare1}d) and in fact are able to recover the imperfections and grain-like features. We underscore that this requires three elements: (a) background chemical patterns, (b) diffusiophoretic motion for a cell in response to chemical gradient, and (c) an interaction potential between cells, modeled here as a hard-sphere potential for simplicity.

\par{} There is evidence supporting that all three elements are present in natural systems. For instance, Kratochwil and Mallarino~\cite{kratochwil2023} argue that in vertebrate color patterns, there are two steps: pre-patterning and patterning. In our model,  chemical patterns resemble the pre-patterning step and diffusiophoretic response of cells resemble the patterning step. Moreover, it is established that entities such as chromatophores and biological cargo are diffusiophoretic~\cite{liu2023, ramm2021diffusiophoretic}, which highlights the importance of diffusiophoresis to be included in a proposed mechanism. Finally, natural entities clearly interact with each other and in fact, colloidal interactions have been argued to be critical for stability of a living cell~\cite{wennerstrom2020colloidal}.
\begin{figure*}[htb]
    \centering
    \includegraphics[width=\textwidth]{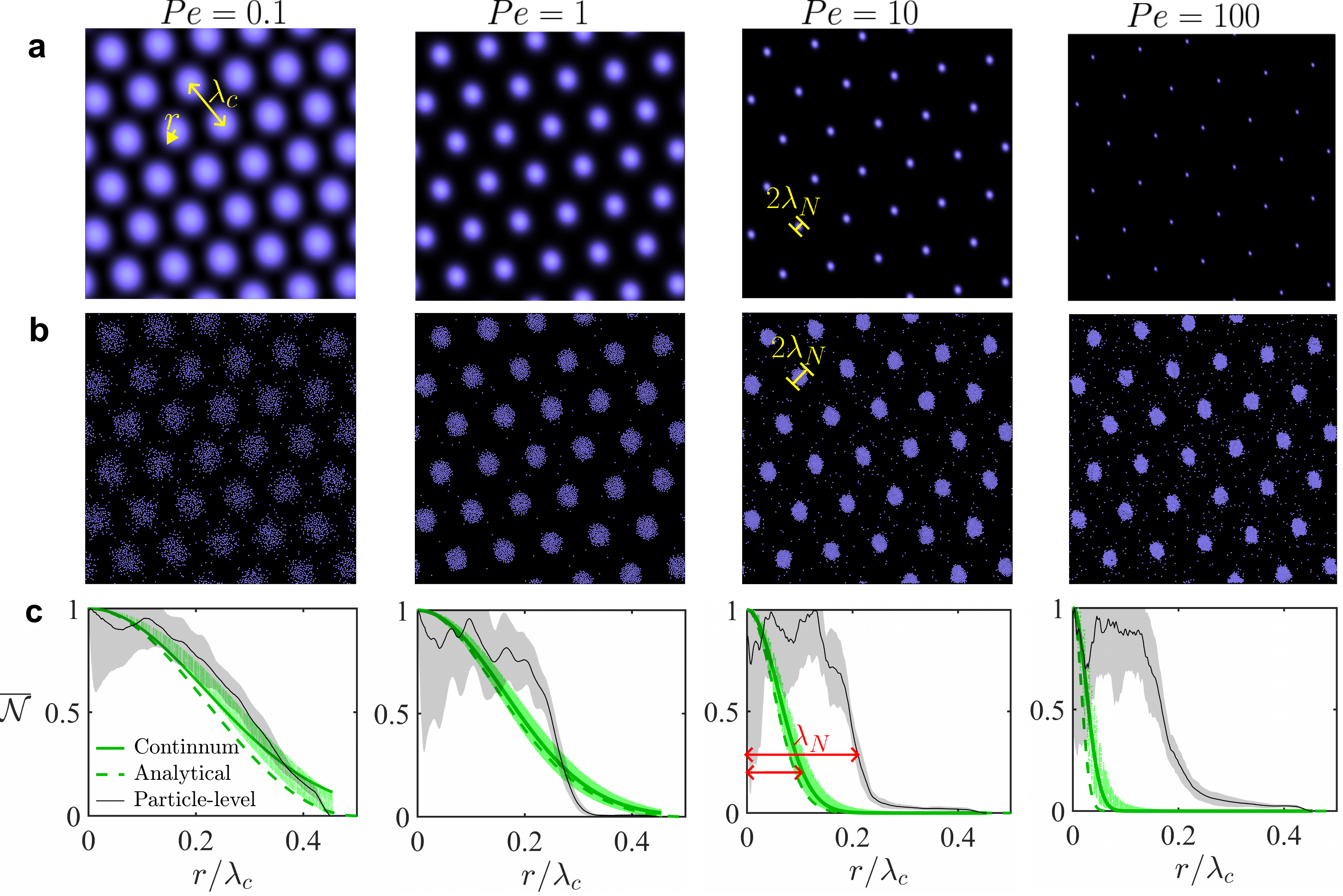}
    \caption{
 \textbf{Comparison between the particle-level and continuum models \SM{in simulating cell assembly}}. \SM{Steady-state cell concentration fields computed from (a) the continuum model and (b) the particle-level simulation for different P\'eclet numbers $Pe$.} \SM{(c) Comparison between the two models in predicting azimuthally averaged cell concentration, $\overline{\mathcal{N}}$, as a function of radial distance $r$ from the center of a hexagon}. \SM{Significant differences emerge between the continuum and particle-level models in the high Péclet number regime ($Pe \gg 1$) where the former predicts a collapse of all particles to the center of the hexagons while the latter yields finite-size clusters.}  Shaded regions represent the standard deviation of the numerical data points. \SM{See the Methods section for details on the derivation of $\overline{\mathcal{N}}$, corresponding error estimates, and the parameters used in the simulations.}
    } 
    \label{fig:compareC}
\end{figure*}

\section*{Results}

We now discuss the details of the proposed framework and how it captures key features observed in natural systems. To simulate a cell pattern, we start with a homogeneous distribution of solutes. These solutes are governed by a set of reaction-diffusion equations 
\begin{equation}\label{eq:RD1}
     \frac{\partial c_i }{\partial \mathfrak{t}} = D_{c_i} {\nabla^2} c_i + R_{c_i},
\end{equation}
where $c_i$ is the concentration of solute molecules and $i$ refers to the particular solute species. $D_{c_i}$ and $R_{c_i}$ represent the diffusion coefficient and production rate of the $i$th solute, respectively. 
\SM{The relative importance of the reaction rate over the solute diffusion is quantified by the dimensionless Damk\"ohler number, defined as $Da_c = kl^2/D_{c_1}$, where $k$ is the equivalent first-order reaction constant and $l$ is a characteristic length scale; see Methods for more details.}
Equation~\eqref{eq:RD1} is solved under periodic boundary conditions and is currently uncoupled from the cell distribution, though it is straightforward to introduce such a coupling~\cite{alessio2023}.  \SM{Such an extension could incorporate interactions between particles and solutes by expressing $R_{c_i}$ as a function of both solute and cell concentrations, as proposed by Nakamasu \textit{et al}.~\cite{nakamasu2009}. In addition, one could account for hydrodynamic disturbances generated by diffusiophoretic motion. These disturbances could then be used to correct the solute conservation equation in equation~\eqref{eq:RD1}. We note that such two-way coupling becomes especially important in concentrated systems.}
\par{} Having obtained $c_i$ from equation~\eqref{eq:RD1}, the diffusiophoretic velocity of the $j$th cell is evaluated to be $\boldsymbol{v}^{dp}_j = \sum_i m_{ji}{\boldsymbol{\nabla}} {c_i}$ \cite{anderson1982}, where $m_{ji}$ is the diffusiophoretic mobility of the $j$th cell with respect to the $i$th solute. 
For cells, we take two different approaches: continuum and particle-level. For the continuum approach, we compute the concentration field of the $j$th cell, $n_j$ by solving the following advection-diffusion equation:
 \begin{equation}\label{eq:AD1}
     \frac{\partial n_j}{\partial \mathfrak{t}} + \boldsymbol{\nabla \cdot} (\boldsymbol{v}^{dp}_j n_j)  = D_{n_j} \nabla^2 n_j, 
 \end{equation}
where $D_{n_j}$ is the diffusivity of the $j$th cell.
\SM{The relative importance of diffusiophoresis and diffusion is controlled by the P\'eclet number, $Pe \approx \frac{mc^*}{D_{n_j}}$, where $m$ denotes a characteristic diffusiophoretic mobility and $c^*$ is the reference solute concentration. A precise mathematical definition of $Pe$  is provided in the Methods section.
%\begin{equation}\label{eq:pe}
%    Pe = \frac{\alpha\left | \mathcal{M}_1  - \mathcal{M}_2 \frac{\eta (1+ A\eta)}{A} \right |}{\mathcal{D}_N},
%\end{equation}
%where $\alpha$ is the perturbation amplitude to $C_1$. We refer the reader to \citep{alessio2023} for more details on the mathematical derivation of $Pe$.
%Definitions of the parameters appearing in equation~\ref{eq:pe} are provided in the Methods section.
} 
For the particle-level approach, we consider a two-dimensional dispersion of $N_p$ rigid particles, where each individual particle is explicitly tracked over time. The particle motion is described by the $N_p$-body Langevin equation as written by
\begin{equation}
     \mathbf{m} \cdot \frac{d\boldsymbol{u}}{d\mathfrak{t}} = \boldsymbol{F}^H + \boldsymbol{F}^P + \boldsymbol{F}^B,
     \label{eq:particle}
\end{equation}
where $\mathbf{m}$ is the generalized mass/moment of inertia $6N_p\times 6N_p$ tensor, $\boldsymbol{u}$ is the translational/rotational velocity vector of the particle of dimension $6N_p$. The total force/torque acting on the particles is the sum of the hydrodynamic forces $\boldsymbol{F}^H$, non-hydrodynamic forces $\boldsymbol{F}^p$, and the stochastic forces $\boldsymbol{F}^B$, which give rise to Brownian motion. Since our cells are small, the inertial term in equation~\eqref{eq:particle} is neglected and the remainder terms are expressed using mobility formulation; see ``Methods" for details. We note that equation~\eqref{eq:AD1} does not capture the steric effects if the cells concentrate near a region. However, in equation~\eqref{eq:particle}, we do capture the effective size as each cell is modeled through its own equation and we also include a hard-sphere interaction potential between the cells.

 We present a systematic comparison between continuum and particle-level approaches in simulating cell assembly for cell sizes significantly smaller than the pattern thickness in Fig.~\ref{fig:compareC}. Here, we employ the Brusselator model, which consists of two solute species, such that it yields chemical patterns in a hexagonal geometry; see ``Methods" for details. To reduce the parameter space, we focus on only a single cell type where the mobility is equal in magnitude with respect to the first and second solute species, but opposite in sign. We observe that in both  approaches, the cell patterns mimic the hexagonal pattern of the solute. However, the details of patterns between these two approaches start to differentiate as a function of P\'eclet number, $Pe$, which is defined as the ratio of cell advection and diffusion; see equation~\eqref{eq:pe} for a mathematical definition. 
\par{} At $Pe = 0.1$ and $Pe=1$, the two approaches converge and yield similar patterns. To quantify the comparison between them, we report the azimuthal average of the concentration of cells over all hexagons, $\overline{\mathcal{N}}$, as a function of distance from the center of a hexagon $r$ (normalized by center-to-center distance, $\lambda_C$); (see the plots at the bottom of Fig.~\ref{fig:compareC}). We also plot analytical solutions \cite{alessio2023, pena2001} obtained from a perturbation analysis for the continuum approach. First, we note that for particle-level simulations, there is distribution of $\overline{\mathcal{N}}$, which reflects the heterogeneity that naturally arises due to the discrete simulations. For $Pe=0.1$ and $Pe=1$, there is quantitative comparison between the continuum and particle-level simulations, although some deviations start to emerge even at $Pe=1$. These deviations become significant for $Pe=10$ and $Pe=100$. For $Pe \gg 1$, the continuum simulations predict a collapse of all the particles to the center of the hexagons. While the particle-level approach also demonstrates clustering, the hard-sphere interactions ensure that a finite cluster size is obtained. We note that the clusters in particle-level simulations are not uniform in shape, signaling the appearance of imperfections which are absent in the continuum approach. 
\begin{figure}[H]
    \centering
    \includegraphics[width=0.49 \textwidth]{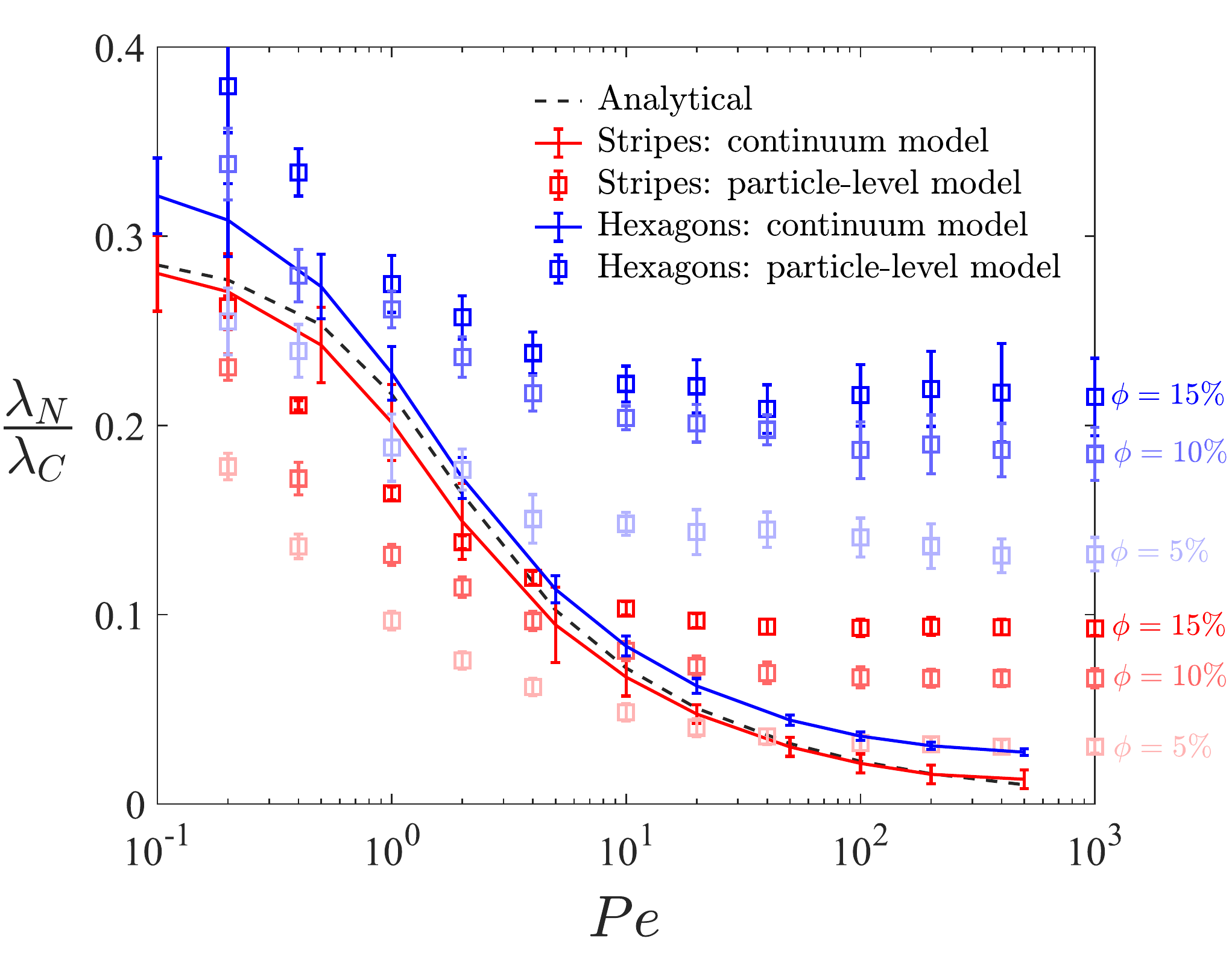}
    \caption{
    \textbf{\SM{Master curve showing dependence of cell pattern thicknesses $\lambda_N$ on the P\'eclet number $Pe$ in the limit $d/\lambda_C \ll 1$}.} Comparison between continuum and particle-level simulation results over a wide range of $Pe$, for both stripe and hexagonal patterns. The prediction from the continuum model is independent of $\phi$, while the particle-level simulations show a notable dependence on $\phi$.  To confirm that the results are within the limit of small cell sizes compared to pattern thickness, we perform simulations with smaller cells and observe quantitatively similar results. Error bars represent the standard deviation of the local pattern thickness. Details of the simulation method and parameters are provided in the Methods section.
    }
    \label{fig:compare2}
\end{figure}
\par{} By utilizing the $\overline{\mathcal{N}}$ versus $r/\lambda_C$ data, we define $\lambda_N$, which evaluate the wavelengths of cell patterns (Fig.~\ref{fig:compareC}). We argue that $\lambda_N/\lambda_C$ is a key parameter for cell patterning as it measures how the cell patterns differ from the chemical patterns. We calculate $\lambda_N/\lambda_C$ for both the hexagon and stripe-like patterns from the Brusselator model, and for both the continuum and particle-level approaches. As previously reported \cite{alessio2023}, continuum model predicts a monotonic decrease of $\lambda_N/\lambda_C$ with increasing $Pe$. We seek to address the question: what happens for the particle-level simulations? This is the focus of Fig.~\ref{fig:compare2}. 
\par{} Several comments are in order regarding the trends reported in Fig.~\ref{fig:compare2}. For small $Pe < O(1)$, we observe that there is a monotonic decay of $\lambda_N/\lambda_C$ with Pe in both the continuum and particle-level models. However, some differences exist. First, the two approaches do not converge to the same value of $\lambda_N/\lambda_C$ because the method used to calculate $\lambda_N$ differs between the two due to the stochastic nature of the particle-level simulations; see ``Methods" for details. Second, the particle-level simulation clearly displays a trend with initial \SM{area} fraction $\phi$. For small $\phi$, the $\lambda_N/\lambda_C$ values tend to be smaller, which indicates sharper cell patterns. This is expected as smaller $\phi$ conditions are less susceptible to saturation due to clustering. 
\begin{figure*}[ht]
    \centering    \includegraphics[width=\textwidth]{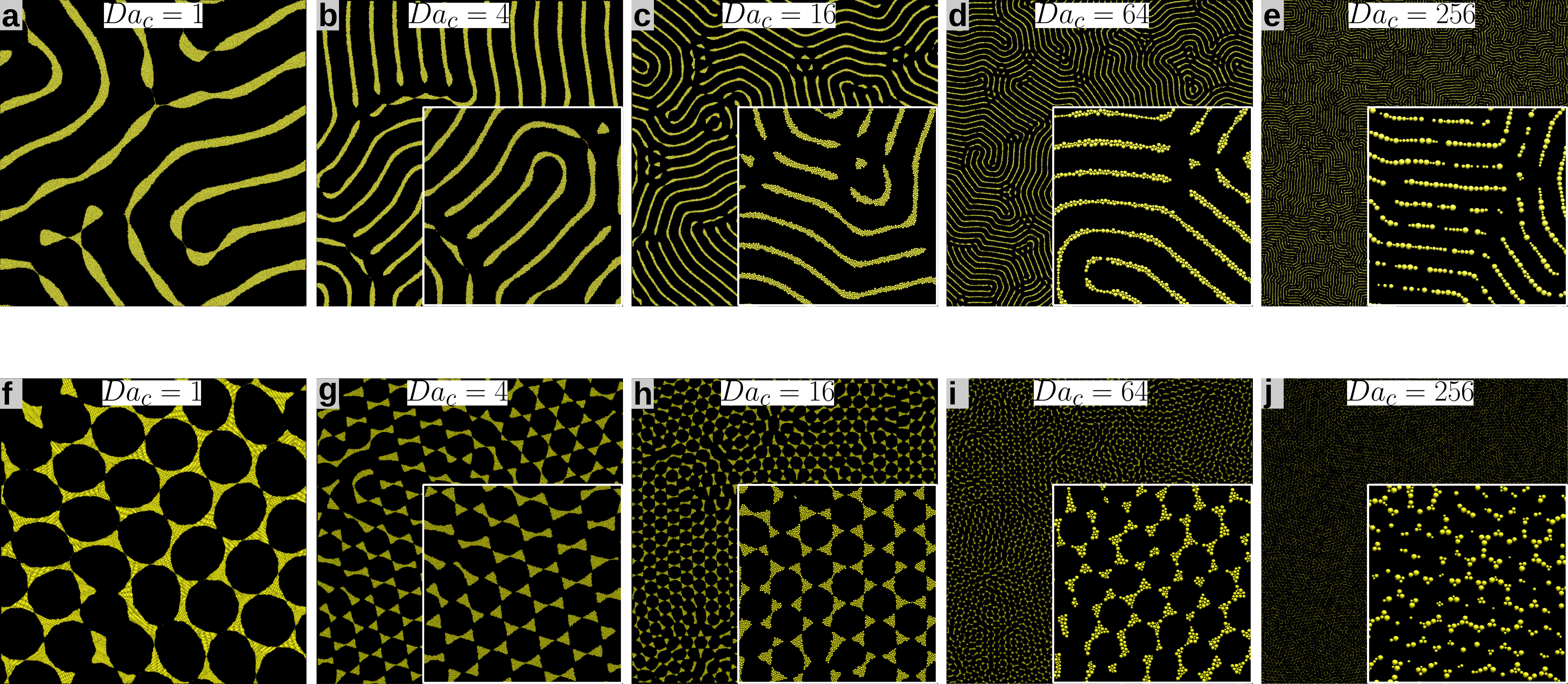}
    \caption{
    \textbf{Impact of cell size on emerging stripe and hexagonal patterns}: Cell assemblies in $\phi = 15\%$ polydisperse systems are examined for different Damk\"{o}hler numbers $Da_c$.  \SM{(a-e) Hexagon and (f-j) stripe pattern type using the same cell size distribution for all simulations with} cell sizes (diameters) randomly assigned within the range $[0.5d, 1.4d]$ where $d$ is the characteristic particle size of the simulation. Note that the pattern size $\lambda_c$ decreases with increasing Damköhler numbers $Da_c$, following \SM{ the scaling relation} $\lambda_C \sim Da_c^{-1/2}$, \SM{thereby reducing the ratio between pattern features and cell size}. Imperfections emerge at higher $Da_c$ as the relative particle size approaches the characteristic thickness of the pattern, eventually preventing the formation of a complete pattern when particles become too large. The same parameters detailed in the Methods are used for solute species, except that the Damköhler number ($Da_c$) is varied.
    }
    \label{fig:snapshot_dac}
\end{figure*}
 For $Pe \gg O(1)$, $\lambda_N/\lambda_C$ continues to decrease with an increase in $Pe$ for continuum simulations. In contrast, for particle simulations, $\lambda_N$ plateaus and no longer steepens with an increase in $Pe$. The plateau occurs because the local packing limit is reached and further increase in diffusiophoretic strength cannot steepen $\lambda_N$. In this limit, it is possible to estimate $\lambda_N/\lambda_C$ from a geometric argument combined with conservation of cells. We refer to the estimation of $\lambda_N$ in this limit as $\lambda_{N, pack}.$ For example, for monodisperse spheres, the densest ordered packing in two-dimensional space is face-centered-cubic (FCC) lattice, corresponding to the area fraction of $\phi_{max} \approx 0.9$ \cite{trovato2007}. Therefore, assuming a circular shape of the cluster at maximum packing, $\pi \lambda_{N,pack}^2 \phi_{max}$ would be equivalent to the total amount of cells. This should be identical to the particles distributed uniformly with \SM{area} fraction $\phi$ in a hexagon with area $\frac{\sqrt{3}}{2} \lambda_C^2$ (recall that edge length is $\frac{2}{\sqrt{3}} \lambda_C$ since $\lambda_C$ is center-center to distance between two hexagons that share an edge), and hence \begin{equation}
       \frac{\lambda_{N,pack}}{\lambda_C} = \left[ \frac{\sqrt{3}}{2 \pi} \frac{\phi}{\phi_{max}} \right]^{\frac{1}{2}}.
       \label{eq:geom1}
\end{equation}
For $\phi=0.15$ and $\phi_{max} \approx 0.9$, $\lambda_{N,pack}/\lambda_C \approx 0.21$, which is in quantitative agreement with the numerical simulations; see Fig.~\ref{fig:compare2}. For a stripe arrangement of cells, the calculation is straightforward: $\lambda_C \phi$ needs to be equated with $\lambda_{N,pack} \phi_{max}$ to get 
\begin{equation}
    \frac{\lambda_{N,pack}}{\lambda_C} = \frac{\phi}{\phi_{max}}.
    \label{eq:geom2}
\end{equation}
This yields that for $\phi=0.15$ and $\phi_{max}=0.9$, $\lambda_{N,pack}/\lambda_C$ = 0.16, which is in reasonable quantitative experiments with the simulations. This analysis also highlights that the behavior of $\lambda_N$ with $\phi$ varies with the type of pattern. While one can estimate the steepening for simple patterns like hexagons and stripes, for more complex patterns~\cite{kondo2010}, simulations are required to predict the details. We note that these calculations are based on a hard-sphere interactions, and other interaction potentials will influence the phase diagram presented in Fig.~\ref{fig:compare2}. 
\par{} Besides the interaction potential, another factor that impacts pattern formation is the relative size of cells. So far, our analysis has been based on the limit of small cells, i.e., when the cell diameters  $d \ll \lambda_N$. Moreover, we have only considered systems with a single particle size. Now, we relax these assumptions and elucidate how they further yield imperfections in Turing patterns. \par{} To investigate the effect of finite cell size, we take a random distribution of sizes which are kept the same across different simulations. We systematically decrease $\lambda_N$ by changing the Damkohler number $Da_c$ of the chemical pattern (see ``Methods" for details), which consequently changes the $\lambda_C$. For a higher $Da_c$, $\lambda_C$ decreases as $\lambda_C \sim Da_c^{-1/2}$. We set $Pe \gg 1$ to ensure we reach the packing limit. Therefore, increasing $Da_c$ allows us to approach the limit where $d/\lambda_N = O(1)$. The results are summarized in Fig.~\ref{fig:snapshot_dac}.

For $Da_c=1$, when $d \ll \lambda_N$, the patterns display spatial heterogeneity that arises from the polydisperse nature of cells, though the patterns are smooth and lack a grain-like feature. As $Da_c$ increases, the patterns start to display grain-like feature where the thickness of the patterns start to change within a repeating unit. At $Da_c = 256$, we observe incomplete pattern formation and high imperfections. This highlights the rich and diverse range of patterns that a particle-level simulation is able to produce by combining diffusiophoresis-assisted assembly and interaction between cells. 
\par
\begin{figure}[H]
    \centering
    \includegraphics[width=0.45 \textwidth]{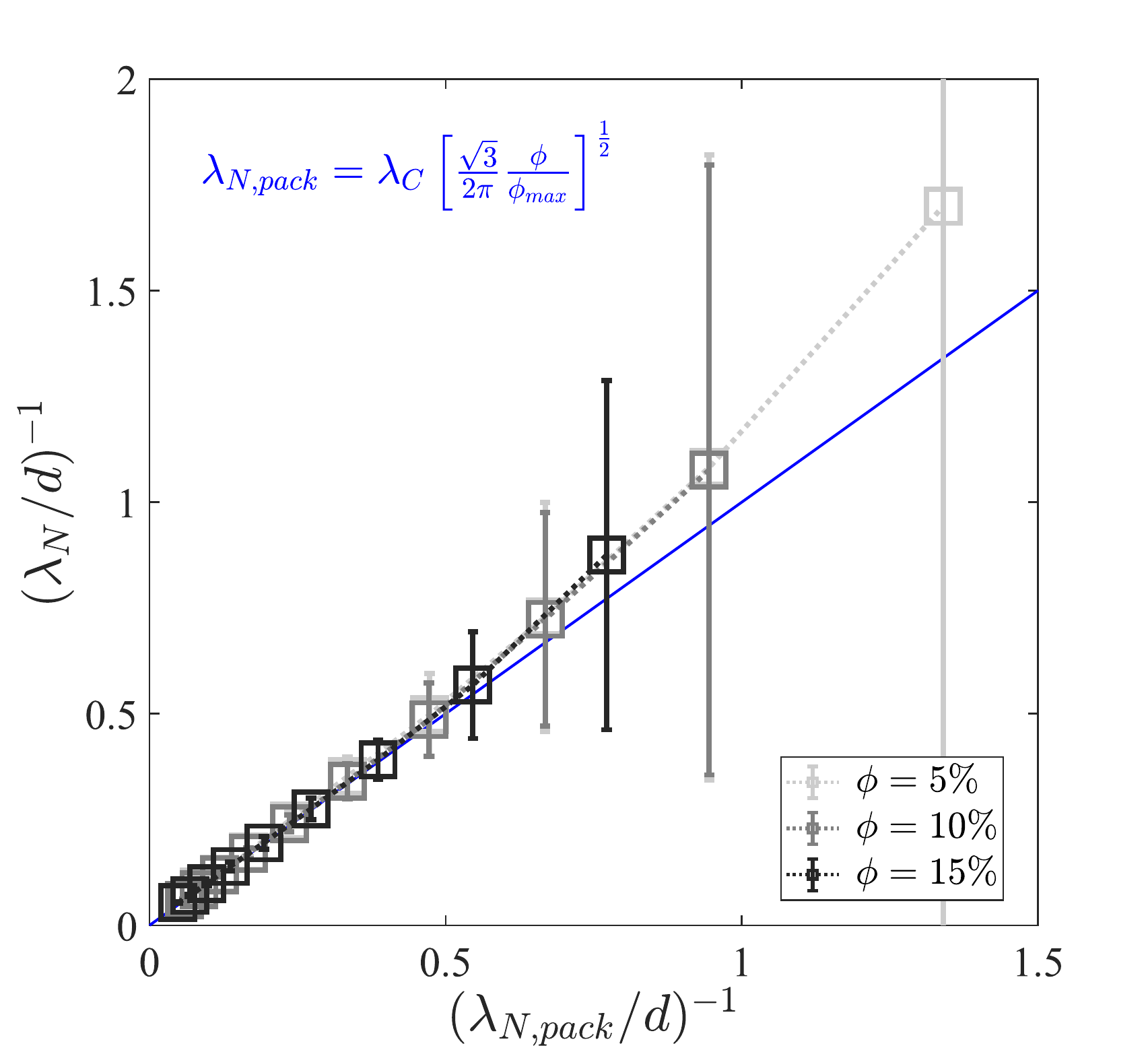}
    \caption{
\textbf{Effect of cell size on \SM{pattern edge thickness} for hexagonal patterns}. \SM{Comparison between the pattern thickness $\lambda_N$ computed by particle-level simulations and the asymptotic prediction $\lambda_{N,\text{pack}}$ (equation (4)) in the high P\'eclet number limit ($Pe = 100$).}
\SM{As the asymptotic prediction $\lambda_{N, pack}$ approaches the particle size $d$, the $\lambda_N$ computed from the particle-level simulations exhibits large variance, signaling notable imperfections and incomplete patterns}. We \SM{set} the same model parameters detailed in Methods to simulate the solute concentrations for the hexagonal pattern \SM{shown in Fig.~\ref{fig:compareC}}. 
    }
    \label{fig:size_snap}
\end{figure}

{} To further quantify the imperfection, Fig.~\ref{fig:size_snap} shows that the edge thickness calculated from the particle simulation ${\lambda_N}$ starts to deviate from the $\lambda_{N,pack}$ predicted by equation~\eqref{eq:geom1}. For simplicity, we only focus on a monodisperse system with a hexagonal pattern. The results show a significant variance in $\lambda_N/d$ when $\lambda_{N,pack}/d > 0.5$.  This happens due to the notable grain-like features which display heterogeneity in pattern thickness. When $\lambda_{N,pack}/d > 1$, extremely high error bars indicated that cells are no longer able to form consistent patterns.

\section*{Discussion}
The results outlined in this article show that the proposed framework -- combining diffusiophoretically-assisted assembly of finite-sized cells that also interact with each other -- opens a new line of inquiry in pattern formation, one which creates patterns that are textured and imperfect, in contrast to classical Turing patterns. Since Turing patterns underpin a range of process~\cite{kondo2010} such as morphogensis, limb development, pattern formation on vertebrate skin and many others, introducing imperfections to them could have profound implications on our understanding. For instance, if the stripes and spots appearing are imperfect, the consequent growth and function could vary significantly. Our analysis shows that these imperfections could be tuned with \SM{area} fraction, particle size, and its distribution, among other factors. While this article focused on the Brusselator model, chemical patterns can utilize a wide range of reaction mechanisms to access a variety of blueprints over which cells can assemble. Furthermore, we only focused on a hard-sphere interaction potential between the cells, which can be readily generalized to other potentials such as Van der Waals and Lennard-Jones, among others~\cite{israelachvili2011intermolecular}. Other factors also impact patterns such as growth and noise and can be explored for particle simulations~\cite{maini2012turing}. There is also an exciting possibility of exploring the 3-dimensional assembly of cells using the same framework. \SM{It should be noted that in actual biological systems, particularly those involving fish skin pigmentation, pigment cells are typically smaller than the thickness of the pigment-containing layer. This allows for vertical as well as lateral arrangement of cells, thereby potentially influencing the effective pattern thickness. In the current study, we reduce the dimensionality from three dimensions (3D) to two dimensions (2D) under the assumption that the vertical thickness is significantly smaller than the lateral dimensions (i.e., $L_z \ll L_x, L_y$) and that the surface topography is relatively flat. This dimensionality reduction saves considerable computational resources. However, future quantitative simulations should explore the exciting possibility of including 3D effects such as cell aggregation and skin topological variation, both of which can significantly influence patterns \cite{manukyan2017, fofonjka2021reaction}.} Furthermore, the inclusion of dissipative interactions can introduce additional local structural features (i.e., specific length scales), such as by contributing to a banding instability in the presence of (shear) stresses \cite{lerouge2006} and the emergence of porosity bands \cite{rudge2024}.

\SM{We acknowledge that while our model generalized the theoretical framework of Turing patterns, the precise biological underpinnings remain only partially understood. For instance, our current theoretical framework, much like the entire field of Turing patterns, simplifies biological mechanisms by ignoring what precise biochemical reactions might be occurring in a real system and by lumping transport challenges that may arise due to cellular or extracellular barriers into cellular mobility. Biological studies have begun to unravel some of these questions. Kratochwil \textit{et al.}~\cite{kratochwil2023} suggest that pattern formation generally involves two sequential processes: the establishment of positional or chemical pre-patterns through morphogen gradients and instructive paracrine factors originating from neighboring tissues or
developing organs, and subsequent pattern implementation via cellular processes. For instance, biochemical signals from somites can establish positional pre-patterns guiding pigment-cell arrangement across multi-cellular distances in galliform birds \cite{haupaix2018}. Furthermore, studies on anole lizards demonstrate how biochemical alterations in reaction–diffusion processes modulate pigment-cell migration, transitioning between distinct color patterns \cite{feiner2022}. We also acknowledge that actual biological patterning my involve additional mechanisms, such as pigment production and cellular interactions,  that our simplified theoretical model does not incorporate. An example includes African cichlid fish, where the migration of chromatophore progenitors along myoseptam and pigment production both contribute to the pattern development \cite{hendrick2019,liang2020,kratochwil2023}.  
Future extensions of this framework could integrate such biological complexities to bridge theoretical predictions and biological systems.}

\SM{We underscore that the current RD-driven diffusiophoresis framework is particularly relevant to biological phase separation, such as phase-separated protein-rich droplets and partitioning of enzymes \cite{testa2021} and dialytaxic swimming of chemically active biomolecular condensates \cite{jambon2024}. Recent findings demonstrate that diffusiophoretic mechanisms can promote phase separation, enhance the motility of condensates, and govern their intracellular localization \cite{doan2024}. Particularly, biochemical reactions driven by chemical fuels and waste production naturally generate concentration gradients that give rise to diffusiophoretic transport of liquid condensates, providing a simple yet effective mechanism for intracellular organization \cite{hafner2024}.}

\par{} From an application standpoint, our work is relevant for a variety of fields where Turing patterns of small-scale entities are of interest. As shown in Fig.~\ref{fig:compare1}, we are able to replicate natural patterns with striking similarity. Expanding this analysis to a richer variety of seed patterns and cell shapes will enhance the fundamental understanding of a variety of patterns that remain poorly understood~\cite{kratochwil2023}. The approach outlined is also important for polyamide membranes where nanoscale Turing structures have been utilized for improved water filtration~\cite{tan2018polyamide}. Since the particles assemble on top of these Turing structures, we believe that the use of diffusiophoresis presents an opportunity to improve the design of better membranes. Specifically, given the ability of our method to obtain grain-like and textured features, we believe this could be employed for the development of colloidal-laden materials where reaction-diffusion instabilities from the dissolved species can help impart anisotropic properties. In particular, leveraging instabilities for controlling morphing dynamics has inspired new designs of programmable soft robotics \cite{jones2021} and this work could inspire the inclusion of diffusiophoresis to impart anisotropy via colloidal assembly in future designs.

\section*{Methods}
For creating Turing patterns in the current study, we particularly consider a four-component system comprising two types of solutes and two types of cells (chromatophores).

The concentrations of solute species are given by the reaction-diffusion equation (equation~\eqref{eq:RD1}), rewritten in nondimensional form as
 \begin{equation}\label{eq:RD2} 
     \frac{\partial C_i }{\partial t} = \mathcal{D}_{C_i} {\nabla^2} C_i + \mathcal{R}_{C_i}, \ \ \ i = 1, 2, 
\end{equation}
where the non-dimensional quantities are introduced by  $C_i = c_i/c^*$, $t = \mathfrak{t}D_{c_i}/l^2$, $\mathcal{D}_{C_i} = D_{c_i}/D_{c_1}$, and $\mathcal{R}_{C_i} = R_{c_i}l^2/c^*D_{c_i}$. Note that $l$ and $c^*$ are reference length scale and reference solute concentration, respectively. Here, we describe the production rate function $\mathcal{R}$ in equation~\eqref{eq:RD2} using the Brusselator model \cite{prigogine1967,prigogine1968} that can be given by
\begin{equation}\label{activator}
    \mathcal{R}_{C_1} = Da_c \left ( A- (B+1)C_1 + C^2_1 C_2 \right )
\end{equation}
\begin{equation}\label{subtrate}
    \mathcal{R}_{C_2} = Da_c \left ( B C_1 - C^2_1 C_2 \right ),
\end{equation}
where $A$, $B$ are input control parameters (see \cite{pena2001} for details). 
%The Damk\"{o}hler number $Da_c = kl^2/D_{c_1}$ is the nondimensional parameter that measures the reaction rate over solute diffusion, where $k$ is the first-order reaction constant. 
We solve equation~\eqref{eq:RD2} by creating finite-difference solvers using two different methods: explicit method based on Adam Bashforth scheme and a semi-implicit method. For all the simulations, a periodic domain with dimensions  $ L \times L = 32l \times 32l$  is used to exclude any boundary effects. We set \SM{the Damk\"{o}hler number} $Da_c = 1$ (except in Figs.~\eqref{fig:snapshot_dac} and \eqref{fig:size_snap}, where $Da_c$ is varied),   $\mathcal{D}_{C_2} = D_{c_2}/D_{c_1} = 8$, and $A = 4.5$. To control pattern morphology, we use the supercriticality parameter defined as $\mu \equiv  B(1+A\eta)^{-2} - 1$ \cite{pena2001} where $\eta = \mathcal{D}^{-1/2}_{C_2}$. Here,  $\mu = 0.04$ and $\mu = 0.1$ are set for hexagons and stripes, respectively. Homogeneous initial conditions are used with $C_1 = A$ and $C_2 = B/A + \xi$ where $\xi$ is the noise uniformly sampled in the range $[-0.01, 0.01]$.

For the continuum representation of the cells, the non-dimensional form of equation~\eqref{eq:AD1}  is given by
\begin{equation} \label{eq:AD2}
     \frac{\partial N_j}{\partial t} + \boldsymbol{\nabla \cdot} (\boldsymbol{V}^{dp}_j N_j)  = \mathcal{D}_{N_j} \nabla^2 N_j, \ \ \ j =1,2.
\end{equation}
Here, $\mathcal{D}_{N_j} = D_{n_j}/D_{c_1}$, $\boldsymbol{V}^{dp}_j = \boldsymbol{v}^{dp}_j l/D_{C_1}$,  $N_j = n_j/n^*$ where $n^*$ is the reference concentration for cells. Both types of cells are considered to have the same diffusivity, i.e., $\mathcal{D}_{N_1} =  \mathcal{D}_{N_2}$. The exact nature of the interaction between the cells and the solute species in relevant biological contexts remains unknown \cite{dukhin2010}. Therefore, for convenience, the diffusiophoretic velocity of the cell species $\boldsymbol{V}^{dp}_j$ is expressed here in a non-electrolyte form as given by
\begin{equation}
    \boldsymbol{V}_j^{dp} = \mathcal{M}_{j1}\boldsymbol{\nabla} C_1 + \mathcal{M}_{j2}\boldsymbol{\nabla}{C_2}, \ \ \ \ j = 1,2.
\end{equation} 
$\mathcal{M}$ is the nondimensional form of the diffusiophoretic mobility. %As inferred from equation~\eqref{eq:AD2}, there are two mechanisms contributing to the cell transport: diffusiophoresis and diffusion. The relative importance of these two mechanisms is controlled by a parameter known as the P\'eclet number (Pe) that is given by
\SM{The P\'eclet number (Pe) that measures the relative importance of the diffusiophoresis and diffusion is given for the current system by}
\begin{equation}\label{eq:pe}
    Pe = \frac{\alpha\left | \mathcal{M}_1  - \mathcal{M}_2 \frac{\eta (1+ A\eta)}{A} \right |}{\mathcal{D}_N},
\end{equation}
where $\alpha$ is the perturbation amplitude to $C_1$. We refer the reader to \citep{alessio2023} for more details on the mathematical derivation of $Pe$.

With the solute concentration gradients obtained from solving the reaction-diffusion description (equation~(\ref{eq:RD2})), the transport equation for cells (equation~\eqref{eq:AD2}) is solved using the finite difference method with an implicit scheme under periodic boundary conditions. A uniform initial condition is applied to both $N_1$ and $N_2$.

A more realistic representation of cell transport is provided by directly treating individual cells using a particle-level model. In this model, we consider the dispersion of spherical cells in a periodic two-dimensional domain. Similar to the continuum approach, two types of cells are considered, each with different diffusiophoretic mobilities. Assuming a dilute dispersion with no hydrodynamic interactions, from the Langevin equation (equation~\eqref{eq:particle}), the total particle translational velocity can be expressed as:
\begin{equation} \label{eq:PS}
     \boldsymbol{U}_i^p = \mathcal{M}_{i1} \boldsymbol{\nabla}C_1 + \mathcal{M}_{i2} \boldsymbol{\nabla} C_2 +  Pe^{-1/2}\Delta t^{-1} \boldsymbol{X}(\Delta t) + \mathbf{R^{-1}_{FU}} \boldsymbol{\cdot F}^p.
\end{equation}
Here, the third term represents the contribution of the Brownian motion, which corresponds to the diffusion part in (equation~\eqref{eq:AD2}). $\boldsymbol{X}(\Delta t)$ is a random displacement that has zero mean $\overline{\boldsymbol{X}} = 0$, and covariance given by $\overline{\boldsymbol{X}(\Delta t)\boldsymbol{X}(\Delta t)} = 2\mathbf{R}^{-1}_{FU}\Delta t$. The resistance tensor for the current model without considering hydrodynamic interactions is given by $\mathbf{R}_{FU} \sim  \mathbf{I}$, where $\mathbf{I}$ is the identity matrix. Here, the P\'{e}clet number is also given by equation~\eqref{eq:pe} with the Brownian diffusion coefficient $D_{br} = M_0 k_B T$ substituted for \SM{$D_{n}$}. Here, $M_0$ represents the mobility of a single cell, defined for a sphere as $(6\pi \eta_0 a)^{-1}$, where $\eta_0$ is the medium viscosity and $a$ is the radius of the sphere. It should be noted that $D_{br}/D_{c_1}\sim \mathcal{D}_N$. 

With the cell velocities calculated by equation~\eqref{eq:PS}, the cell positions are updated over time with a second-order Adams-Bashforth time-marching scheme with an explicit Euler scheme employed for the initial time step. For a fast simulation that handles self-assembly and short-range interactions between many particles/cells (e.g., of $O(10^6)$),  a contact detecting algorithm is developed based on subdomain search, where the particle overlap is prevented using a potential-free algorithm \cite{melrose1993,mirfendereski2022}.
\SM{For the particle-level simulations, the characteristic particle diameter is set to $d = 0.002L$ except for Figs.~\ref{fig:compareC} and \ref{fig:compare2}, where $d = 0.005L$ is used.}
%For the simulation results presented in Fig.~\ref{fig:compare1} and \ref{fig:snapshot_dac}, we set particle size (diameter) $d=0.002L$. The same $d = 0.0002$ is used as the characteristic particle size for Fig.~\ref{fig:size_snap}. For the Fig.~\ref{fig:compareC} and \ref{fig:compare2}. Note the  the diameters of particles set for the simulation shown}

From the concentration field of solutes derived from solving the equation~\eqref{eq:RD2}, the pattern size $\lambda_c$ can be given by the dominant wavelength, which is computed from the Fourier transform of the spatial concentration field of second solute species $C_2(x,y)$, as written by

\begin{equation}
    \lambda_c = \left ( k_{x,max}^2 + k_{y,max}^2    \right )^{-1/2},
\end{equation} 
where dominant corresponding wave numbers are given by 
\begin{equation}
(k_{x, \text{max}}, k_{y, \text{max}}) = \arg\max_{k_x, k_y} \left| \hat{C}_2(k_x, k_y) \right|.
\end{equation}
$\hat{C}_2(k_x, k_y)$ is the two-dimensional Fourier transform of $C_2$, expressed as:
\begin{equation}
    \hat{C}_2(k_x, k_y) = \int_{-\infty}^{\infty} \int_{-\infty}^{\infty} C_2(x,y) \, e^{-i (k_x x + k_y y)} \, dx \, dy.
\end{equation}
For $Da_c =  1$, the pattern size $\lambda_c$ calculated by this approach gives the  same value as critical chemical wave length $ = 2\pi/\sqrt{A \eta}$. 

To determine the pattern thickness in Figs.~~\ref{fig:compareC}, \ref{fig:compare2} and \ref{fig:size_snap}, $\lambda_N$, we specifically focus on the patterns emerging from a single cell type where the mobilities with respect to the first and second solute species are equal in magnitude but opposite in sign with $-\mathcal{M}_1=\mathcal{M}_2 = \mathcal{M}_0$, where $\mathcal{M}_0>0$. \SM{For Fig.~\ref{fig:snapshot_dac}, mobilities are set similarly but with $\mathcal{M}_0<0$.}  We first identify the local maxima in $C_2$ denoted as $\mathcal{L} =\{(x_i,y_i)\}$. For striped patterns, these correspond to the finely-resolved ridge points along the stripes, while for hexagonal patterns, they correspond to the maximum point at the center of each hexagon. 

For the continuum simulation, the average concentration as a function of distance from the maximum concentration point (either ridges or hexagon centers) is defined as
\begin{equation}
    {\mathcal{N}} (r) = \frac{1}{\mathcal{A}(r)} \iint\limits_{r(x, y) \in [r, r + d r]} N(x, y) \, dx\, dy.
\end{equation}
Here, $\mathcal{A}(r) = \iint\limits_{r(x, y) \in [r, r + d r]} dx\, dy$, is the area of the region where the distance to the nearest ridge satisfies  $r(x, y) \in [r, r + d r]$. The distance $r(x,y)$ from any point $(x,y)$ to the nearest ridge point (or center of a hexagon) is given by
\begin{equation}
    r(x, y) = \min_{(x_i, y_i) \in \mathcal{L}} \sqrt{(x - x_i)^2 + (y - y_i)^2}.    
\end{equation}
We then calculate the relative concentration by
\begin{equation}
   \overline{\mathcal{N}} (r) = \frac{\mathcal{N}(r) - \min{\mathcal{N}(r)}}{\max{\mathcal{N}(r)} - \min{\mathcal{N}(r)}}.
\end{equation}
The width of the Gaussian fit to $\overline{\mathcal{N}}(r)$ gives $\lambda_N$. Specifically, $\overline{\mathcal{N}}(r)$ is approximated by the Gaussian function $\overline{\mathcal{N}}(r) \approx \mathcal{G}(r) = \mathcal{G}_0 \exp\left( -\frac{r^2}{2c^2} \right)$, where $ \mathcal{G}_0 = 1 $ , and $ c $ gives $\lambda_N$. To quantify the local variability (or uncertainty) - represented by the green shaded error in the bottom plots of Fig.~\ref{fig:compareC} - the standard deviation of the fit is calculated as follows:
\begin{equation}
\sigma_{\overline{\mathcal{N}}}(r) = \sqrt{\frac{1}{\mathcal{A}(r)} \iint\limits_{r(x,y) \in [r, r+dr]} \left( \overline{N}(x,y) - \overline{\mathcal{N}}(r) \right)^2 \, dx\,dy },
\end{equation}
where $\overline{N}(x,y) = (N - \min{N})/(\max{N}-\min{N})$.
For the distribution of discrete finite-sized particles, we can calculate ${\mathcal{N}}$ by
\begin{equation}
    {\mathcal{N}}(r) = {\sum_{k=1}^{N_p}\chi \left( r;r_k \right)},
\end{equation}
where $r_k = \min_{(x_i, y_i) \in \mathcal{L}} \sqrt{(x_k - x_i)^2 + (y_k - y_i)^2}$ is the distance from particle $k$ located at $(x_k,y_k)$ to the nearest point in $\mathcal{L}$. The weighting function, $\chi(r;r_k)$  for stripe patterns is defined as follows:
\begin{equation}
   \chi(r; r_k) =   \begin{cases}
  \displaystyle \sqrt{1- 2 \left( \frac{r-r_k}{d_k} \right )}, & \text{if } |r_k - r| \le \frac{d_k}{2},\\[1ex]
  0, & \text{otherwise.}
  \end{cases}
\end{equation}
Here, $d_k$ is the diameter of particle $k$. For the hexagonal patterns, the weighting function $\chi(r; r_k)$  can be expressed as follows:
\begin{equation}
    \begin{cases}
  \displaystyle \mathcal{H}(\delta) + \mathcal{H}(-\delta)\frac{1}{\pi}\arccos{\left(\frac{r_k^2 + r^2 - \frac{{d_k}^2}{4}}{2rr_k}\right)}, & \text{if } |r_k - 1| \le \frac{d_k}{2},\\[1ex]
  0, & \text{otherwise,}
  \end{cases}
\end{equation}
where $\mathcal{H}$ is the Heaviside function and $\delta = d_k/2 - (r+r_k)$. Here, the pattern thickness  $\lambda_N$ is also given by the width of the Gaussian fit to  $\overline{\mathcal{N}}$.  

For the particle-level simulations, we determine the gray shaded error in the bottom plots of Fig.~\ref{fig:compareC}, by

\begin{equation}
    \sigma_{\overline{\mathcal{N}}}(r) = \sqrt{\frac{1}{|\mathcal{L}|} \sum_{j=1}^{|\mathcal{L}|} \Bigl(\overline{\mathcal{N}}_j(r) - \frac{1}{|\mathcal{L}|} \sum_{j=1}^{|\mathcal{L}|} \overline{\mathcal{N}}_j(r)\Bigr)^2},
\end{equation}
where $\overline{\mathcal{N}}_j$ is the relative concentration for each hexagon.

\section*{Resource Availability}

%\subsection*{Lead contact}
%Requests for further information and resources should be directed to and will 
%be fulfilled by the lead contact, Ankur Gupta (ankur.gupta@colorado.edu)

\subsection*{Materials availability}
This study did not generate new unique materials.

\subsection*{Data and code availability}
A FORTRAN code was developed to perform all simulations, and MATLAB scripts were written for data analysis. All additional information required to reproduce the data reported in this paper is available from the lead contact upon request. The simulation code used for this study is available on Zenodo: \href{https://doi.org/10.5281/zenodo.17109281}{/10.5281/zenodo.17109281} (password: \texttt{Fast\_PLDP@SM2025}).

\section*{Supplemental Videos}
\textbf{Movie S1}: A monodisperse mixture of approximately 60,000 two-type cells undergoes diffusiophoretic transport in response to the chemical blueprint pattern leading to cell self-assembly and formation of a hexagonal pattern. The chemical pattern is given by the Brusselator reaction model with $\mu = 0.04$. \\
\\
\textbf{Movie S2}: A monodisperse mixture of approximately 60,000 two-type cells undergoes diffusiophoretic transport in response to the chemical blueprint pattern leading to cell self-assembly and formation of a stripe pattern. The chemical pattern is given by  the Brusselator reaction model with $\mu = 0.1$.\\
\\
\textbf{Movie S3}: A polydisperse mixture of 180,000 two-type cells undergoes diffusiophoretic transport in response to the chemical blueprint pattern leading to cell self-assembly and formation of a hexagonal pattern. The chemical pattern is given by the Brusselator reaction model with $\mu = 0.04$. \\
\\
\textbf{Movie S4}: A polydisperse mixture of 180,000 two-type cells undergoes diffusiophoretic transport in response to the chemical blueprint pattern leading to cell self-assembly and formation of a stripe pattern. The chemical pattern is given by the Brusselator reaction model with $\mu = 0.1$.

\section*{Acknowledgements}

We thank the NSF (CBET-2238412) CAREER and AFOSR (FA9550-25-1-0176) YIP award for financial support.

\section*{Authors Contributions}
Conceptualization, S.M. and A.G.; investigation, S.M. and A.G.; methodology, S.M. and A.G.; software, S.M.; visualization, S.M.; project administration, A.G.; funding acquisition, A.G.; resources, A.G.; writing – original draft, S.M.; writing – review \& editing, S.M. and A.G.; supervision, A.G.

\section*{Declaration of Interests}
The authors declare no competing interests

\bibliography{ref}

\end{multicols*}
\end{document}